\documentclass{jps-cp}
\usepackage{csquotes}
\usepackage{graphicx}
\usepackage{siunitx}
\usepackage{txfonts} 

\title{Electric Dipole Moment Measurements at Storage Rings}

\author{J\"org \textsc{Pretz}$^{1,2}$ for the JEDI and CPEDM collaborations\\
  \vspace{2mm}
Proceedings 24th International Spin Symposium (18-22 October 2021)}

\inst{$^{1}$Forschungszentrum J\"ulich \\
$^{2}$ RWTH Aachen University, Forschungszentrum J\"ulich }
\email{pretz@physik.rwth-aachen.de}

\recdate{February 14, 2022}

\abst{
Electric Dipole Moments (EDMs) of subatomic particles,  are
considered as one of the most powerful tools to study CP-violation beyond the Standard Model.
Such CP-violating mechanisms are searched for to 
explain the dominance of matter over anti-matter in our universe.
This paper discusses EDM searches of charged hadrons in storage rings.

The document focuses on activities at the existing storage ring COSY at Forschungszentrum J\"ulich, Germany
and the design of a 100\,m circumference prototype ring able to demonstrate
key technologies and components. These include simultaneous clockwise and counter-clockwise beam operation   
with electrostatic bending elements and, by adding a magnetic field, the frozen spin technique.
}

\kword{electric dipole moment, CP-violation, storage ring, axion searches}

\begin{document}
\maketitle

\section{Introduction}

The existence of electric dipole moments (EDMs) of subatomic particles 
(e.g. atoms, certain molecules, hadrons) is only possible
if parity ($P$) and time reversal ($T$) symmetry are violated.
Assuming that the $CPT$ theorem holds, $T$-violation is equivalent to
$CP$-violation.
Note that in this context we talk only about {\em permanent} EDMs.
The well known EDMs of certain molecules
(e.g. H$_2$0, NH$_3$) are not of this nature and don't require
violation of $P$ and $T$ symmetries.
These molecules appear to have a permanent EDM because
of two almost degenerated energy levels of opposite parity.
This implies that the energy levels grow linearly with an applied electric
field -- a sign of a permanent EDM.
However, in very small electric fields $E$,
the energy levels grow quadratically with the electric field strength (quadratic Stark effect).     
This is the case if the interaction energy $eE$, $e$ being the elementary charge, is smaller than
the energy difference of the two almost degenerated energy levels. 
A more detailed discussion can be found in reference~\cite{Wirzba:2016saz}.

\begin{figure}[tbh]
  \centering
\includegraphics[width=0.8\textwidth]{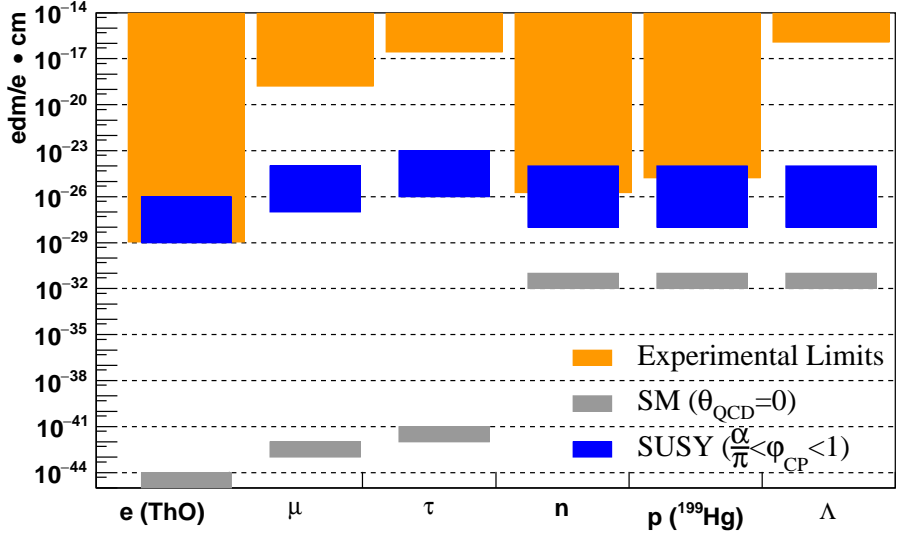}
\caption{90\% EDM limits for various particles, together with prediction from
the Standard Model and Super Symmetry.
\label{fig:edms}}
\end{figure}

The search for EDMs has a long history.
Starting 60 years ago with the measurement of the neutron EDM by
Smith, Purcell and Ramsey~\cite{Smith:1957ht}.
Figure~\ref{fig:edms} shows an overview of experimental results.
Up to now all measurements show results consistent with zero.
The resulting upper limits for various particles (lower edge of orange bar)
together with predictions from super-symmetric models (SUSY) and the Standard
Model are shown. Most of the measurements were performed on neutral systems.
The proton limit was deduced from an EDM measurement  of the mercury atom for example.
One exception is the muon. The limit shown in figure~\ref{fig:edms} was obtained 
at a storage ring experiment where the main purpose was to measure the
anomalous magnetic moment of the muon~\cite{Bennett:2008dy}. 

Based on the principle of the muon EDM measurement, experiments are proposed to measure EDMs of charged
particles in storage rings~\cite{Anastassopoulos:2015ura}. 
The principle will be discussed in the next section.
Section~\ref{cosy} describes measurements at the existing magnetic storage ring
Cooloer Synchrotron COSY at Forschungszentrum J\"ulich, Germany. Section~\ref{ptr} discusses the next step, i.e. plans for a dedicated storage ring to measure EDMs of charged particles.

\section{Principle of Storage Ring EDM Experiments} 
\begin{figure}[tbh]
  \centering
\includegraphics[width=0.8\textwidth]{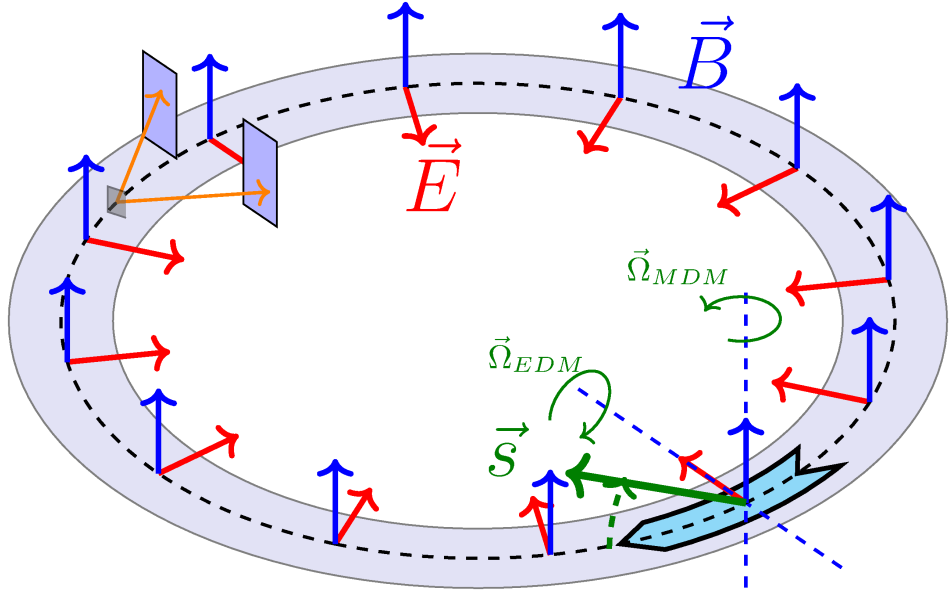}
\caption{Principle of a storage ring EDM measurement.}
\label{fig:princ_mdm_edm}
\end{figure}

For an elementary particle, the spin is the only vector defining a direction.
A permanent EDM has to be aligned along this axis.
If an EDM exists, the spin vector will experience a torque
in addition to the one caused by the magnetic moment. 
For a polarization direction aligned along the momentum vector, this torque causes a polarization component in the
vertical direction.
The polarization direction can be determined by scattering the beam off a carbon target
and analyzing the azimuthal distribution of the scattered particles.
Expected EDMs are of the order of $10^{-27}e\,\si{cm}$. This means that the spin precession
due to an EDM is orders of magnitude smaller than the precession caused by the magnetic dipole moment (MDM).
The magnetic moment causes a precession of the spins
in the horizontal plane as indicated in Fig.~\ref{fig:princ_mdm_edm}.

Quantitatively the spin motion with respect to the cyclotron motion is described by the Thomas-BMT
equation~\cite{Bargmann:1959gz,Nelson:1959zz,Fukuyama:2013ioa}.
Assuming $\vec \beta \cdot \vec E = \vec \beta \cdot \vec B =0$, it is given by
\begin{equation}\label{eq:bmt}
\frac{{\rm d} \vec{s}}{{\rm d} t} = (\vec{\Omega}_{\mathrm{MDM}} + \vec{\Omega}_{\mathrm{EDM}}) \times \vec s \, .
= \frac{-q}{m} 
\left[
  {G} {\vec B}
  +    
\left( G
  -\frac{1}{\gamma^2-1}  \right) \vec v \times \vec  +
  \frac{\eta}{2} 
\large( 
\vec E 
 +
  {\vec v \times \vec B} \large) 
\right]  
\times \vec s
\end{equation}
The variables are explained in Tab.~\ref{tab:var1}.

The vertical polarization component will  oscillate with
a small  amplitude $\beta \eta/(2 G)$ and a angular frequency
$\Omega_{\mathrm{MDM}} \approx - qGB/m$ in a magnetic storage ring.   
This oscillation was used in in the muon $g-2$ experiment~\cite{Bennett:2008dy}
to set a limit on the muon EDM, parameterized by the dimensionless variable $\eta$.
For leptons with $G$-values of approximately $10^{-3}$ this amplitude is three orders of magnitude
larger compared to hadrons where $G \approx \mathcal{O}(1)$.
Therefore, other ways are being sought to obtain a larger signal.

\begin{table}[tbh]
\caption{Definition of variables used in the text. \label{tab:var1}}
\label{t1}
\begin{tabular}{ll}
  \hline
  $s$  & spin in the particle rest frame \\
  $t$  & time in the laboratory frame \\
  $q$ & electric charge \\
  $d = \eta \frac{q \hbar}{2 m c} \vec s$  &  electric dipole moment  \\
  $\eta $     & dimensionless parameter describing the strength of the EDM \\
  $\vec \mu = 2(G+1) \frac{q \hbar}{2m} \vec s$ & magnetic dipole moment \\
  $G$    & anomalous magnetic moment \\
  &  $=-0.1425617662(22) $ for deuterons \\
  &  $=\,\, \, \,1.7928473447(8)   $ for protons \\
  &  $=\,\, \,\, 0.00116592061(41) \approx \frac{\alpha}{2\pi} $ for muons \\    
  $c$  & speed of light \\
  $m$  & particle mass \\
  $\vec E,\vec B$ & electric and magnetic field in the laboratory frame\\
  $\hbar$ & Planck constant/(2$\pi$) \\
  $\beta$ & velocity in units of speed of light $c$ \\
  $\gamma =\frac{1}{\sqrt{1-\beta^2}}$ & \\
  \hline
\end{tabular}
\end{table}

\begin{figure}[tbh]
  \centering
\includegraphics[width=0.8\textwidth]{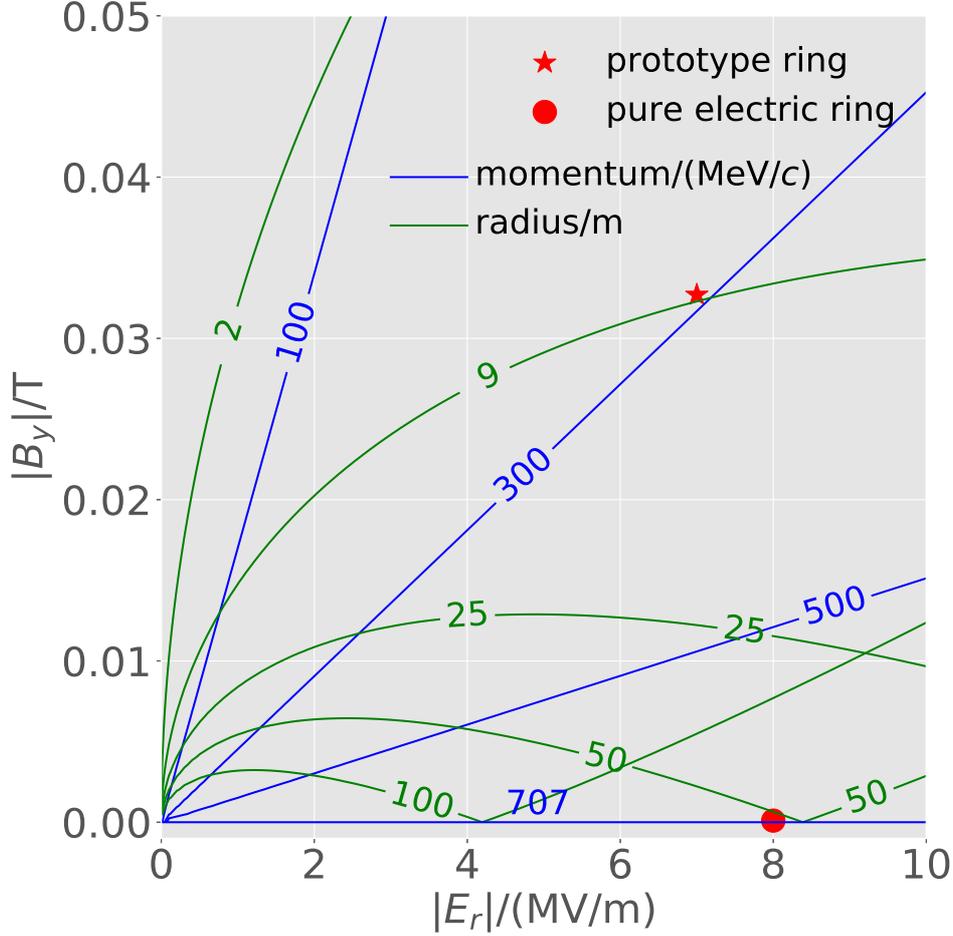}
\caption{Bending radius and momentum as a function of the electric and magnetic bending fields
  for the frozen spin condition. 
\label{fig:p_r_proton}}
\end{figure}

Several methods are proposed. 
One solution is to run in a so called
frozen-spin condition where the precession in the horizontal plane
is suppressed ($\vec \Omega_{\mathrm{MDM}}=0$)
by a suitable electric and magnetic field
combination~\cite{Anastassopoulos:2015ura}.
Possible combinations of electric and magnetic fields leading to
such a frozen spin condition are shown in figure~\ref{fig:p_r_proton}.
In this case the vertical polarization will grow to a larger amplitude
and not just perform a fast oscillation.
In a second method,
which can be applied at a pure magnetic storage ring, 
the horizontal spin precession is influenced by an additional element in the storage ring,
a radio frequency Wien filter in such a way that a build-up due to 
the EDM can also be observed~\cite{PhysRevSTAB.16.114001,Rathmann:2013rqa}.

For the first method a dedicated storage ring has yet to be designed
and built.
The second option can be performed at an existing magnetic storage ring
COSY.
In the next section first deuteron EDM measurements at COSY will be discussed.

\section{Measurements at COSY}\label{cosy}

\subsection{Deuteron EDM measurement}

Two successful deuteron electric dipole moment (EDM) runs have been performed at the Cooler Synchrotron COSY
in December 2018 and March 2021.
Goal of the measurement was the determination of the so-called invariant spin axis $\hat n$,
which is directly related to the EDM. In an ideal accelerator, the invariant spin axis points
in vertical direction along the magnetic guiding field, if the particle has only a magnetic dipole moment and no EDM
($\hat n \parallel \vec \Omega_{\mathrm MDM}$).
The invariant spin axis defines the spin rotation axis. The component of the polarization vector
along the invariant spin axis does not rotate.
The presence of an EDM tilts this axis towards the radial direction by an angle $\eta \beta/(2 G)$,
i.e. $\hat n \propto \vec \Omega_{\mathrm {MDM}} + \vec \Omega_{\mathrm{EDM}} $.
The experiment has been performed with the radio frequency Wien filter~\cite{Slim:2016pim} causing a build-up of
a vertical polarization component, if the invariant spin axis is tilted with respect to magnetic field axis
of the Wien filter. Rotating the Wien filter around the beam axis and by adding a solenoidal field the invariant
spin axis could be deliberately tilted in radial and longitudinal direction, respectively.
Measuring the polarization build-up (in accelerator terminology this means measuring
a resonance strength $\epsilon$)  as a function of both rotations allows one to measure
a two dimensional map, where the minimum indicates the position of the invariant spin axis with the
Wien filter in its nominal position and zero solenoidal field.

\begin{figure}[tbh]
  \centering
\includegraphics[width=0.7\textwidth]{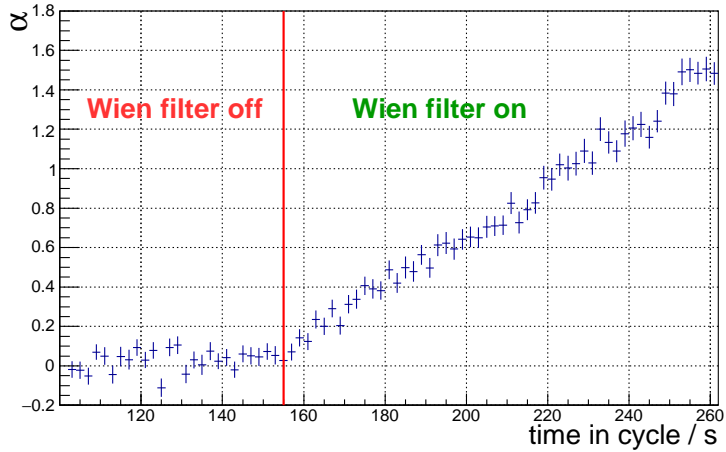}
\caption{Ratio $\alpha = \mbox{arctan}(P_v/P_h)$ of vertical to horizontal polarization component
as a function of time. The time derivative of the slope is directly related to the resonance strength $\epsilon$. 
\label{fig:alpha_2}}
\end{figure}

\begin{figure}[tbh]
    \centering
\includegraphics[width=0.7\textwidth]{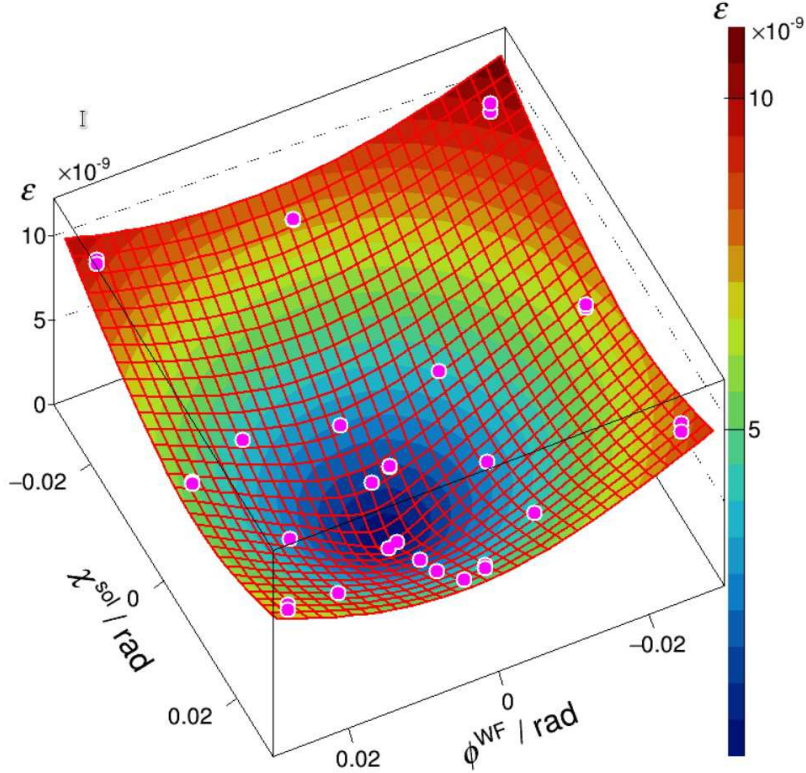}
\caption{Measured resonance strength $\epsilon$ (points) as a function
  of the Wien filter rotation angle $\Phi^{WF}$ and the rotation (spin kick) $\chi^{sol}$ caused by the solenoidal field.
  The colored area shows a fit with a two dimensional parboloid to the data.
The minimum of the map indicates the position of the invariant spin axis.
}
\label{fig:map123}
\end{figure}

Figure~\ref{fig:alpha_2} shows the development of $\alpha=\mbox{atan}(P_v/P_h)$ as a function of time
for a fixed setting of the Wien filter and the solenoid.
Here, $P_{v(h)}$ denotes the vertical (horizontal) polarization component.
As soon as the Wien filter is switched on, $\alpha$, i.e. the vertical polarization
starts to raise. The slope is proportional to the resonance strength $\epsilon$.
Figure~\ref{fig:map123} shows the resonance strength as a function of the
the Wien filter rotation angle $\Phi^{WF}$ and the rotation $\chi^{sol}$ caused by the solenoidal field.
The minimum of the map indicates the position of the invariant spin axis
for the Wien filter in its nominal position and no solenoidal field.
As mentioned above, in an ideal ring one would only expect a tilt of the invariant spin axis in radial direction due to an EDM.
However, here we observe both longitudinal and radial direction tilts of the order of a few mrad,
caused by systematic effects (e.g. misalignments of elements) which are currently under
investigation using beam and spin tracking simulations.
Note that a tilt in radial direction of 1mrad is equivalent to an EDM of $10^{-17} e$cm.

The most important parameter of the experiment are given
in Tab.~\ref{tab:para}.
More details are discussed in the conference contributions~\cite{vitz,saleev,wagner,slim}.


\begin{table}
  \caption{Parameters of the COSY experiment to measure the deuteron EDM.\label{tab:para}}
  \begin{center}
\begin{tabular}{ll}
\hline
COSY circumference  &  183\,m \\
deuteron momentum $p$       &  $\SI{0.970}{\,GeV\per c}$ \\
rel. velocity  $\beta$     & 0.459 \\
Lorentz factor $\gamma$ &   1.126 \\
revolution frequency $f_{\mathrm{cosy}}$ & 750.6\, kHz\\
spin precession frequency $|f_{\mathrm{s}}|$ & 120.9 \,kHz\\
nb. of stored particle  & $\approx 10^{9}$ \\
Wien filter resonance frequency $|f_{\mathrm{rf}}|$ & 873\, kHz\\
Wien filter magnetic field  &  ${2\cdot 10^{-6}}\si{Tm}$ \\
\hline
\end{tabular}
\end{center}
\end{table}

\subsection{Searches for axions and axion-like-particle (ALPs)}
An axion or an axion like particle (ALP) induces an oscillating
EDM, where the oscillation frequency is directly proportional to
the axion/ALP mass $m_a$:
\begin{equation}\label{eq:oscEDM}
  d = d_0 + d_1 \cos(\omega_a t + \varphi_a) \quad \mbox{with} \quad  \omega_a = m_ac^2/\hbar \, . 
  \end{equation}
  If $d_1 \ne 0$ a resonance resulting in a build-up of a vertical polarization
  occurs. The resonance condition is $\omega_a = \Omega_{\mathrm{MDM}}$.
  $\Omega_{\mathrm{MDM}}$ is given by $\gamma G$ in a pure magnetic ring (see eq.~\ref{eq:bmt}). Thus be scanning $\gamma$, axion
  searches for different possible axion masses can be performed.
  The JEDI collaboration scanned in a limited range corresponding a mass range from 4.95 to 5.02 $\times 10^{-9} e$V.
  The big advantage of this method is that in principle one can search at a given mass by running
  at the corresponding frequency.
  One difficulty one has to overcome is the fact that the phase $\varphi_a$ of the axion field with respect
  to the spin precession phase is not known. For this reason four bunches are stored in the accelerator
  with relative polarisation phases of about $\pi/2$.
  More details on the analysis can be found in reference~\cite{karanth:ipac2021}.

\begin{figure}[tbh]
    \centering
  \includegraphics[width=0.7\textwidth]{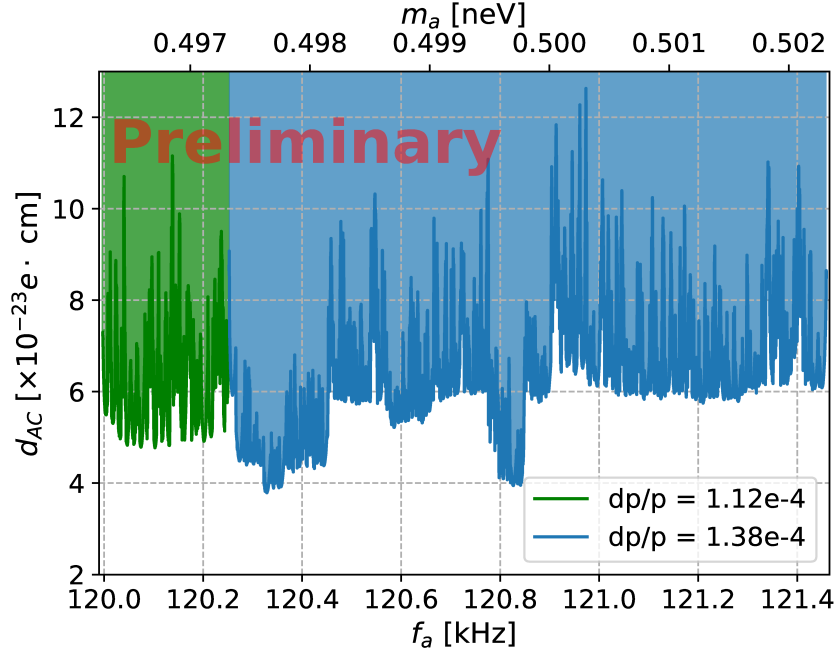}
  \caption{
Preliminary 90\% upper confidence level sensitivity
for an oscillating EDM ($d_1$ in equation \ref{eq:oscEDM}, $d_{AC}$ in the figure)
in the frequency range from 120.0 to 121.4 kHz
(mass = $4.95 - 5.02 \times 10^{-9} e$V).
The green and blue colors show two scanning ramp rates in momentum
change. 
   \label{fig:axion_sen}}
\end{figure}

  No signal was observed. Preliminary results for the resulting 90\% confidence
  limits based on the statistical sensitivity are shown in figure~\ref{fig:axion_sen}. 
  The sensitivity interpreted as an oscillating EDM $d_1$ corresponds to approximately
  to $10^{-22}e$ cm for $d_1$.
  Compared to the search of a permanent EDM (the parameter $d_0$ in equation~\ref{eq:oscEDM})
  the search for an AC effect $d_1$ is less affected by systematic effects. 
  Note that influence  on the spin motion due the axion wind effect is expected to be orders of magnitudes larger \cite{Silenko:2021qgc}
  in storage rings due to the larger velocity ($\beta \approx 1/2$) compared to
  MRT experiments. This effect is still being studied for the current experiment.
  
\section{Design of a dedicated storage ring}\label{ptr}
From the studies at COSY it is evident, that a dedicated storage ring is needed to get a better handle on systematic effects. The most important remedy to fight systematics is the possibility to operate clockwise and counter clockwise beams in the so-called frozen spin mode, where the spin precession due to the magnetic moment is suppressed.

This can be achieved with certain combinations
of electric and magnetic fields.
Figure~\ref{fig:p_r_proton} shows the \enquote{playground} for constructing
a storage where the frozen spin condition ($\vec \Omega_{\mathrm{MDM}}=0$) is fulfilled for protons.
Indicated by the red cicle is an option using a pure electric storage
ring. The main advantage is that two proton beams can run
clockwise and anti-clockwise simultaneously in the ring.
The disadvantage is that, assuming that one is able to operate electrostatic
bends with an electric field strength of $E=\SI{8}{MeV/m}$, the bending
radius of the ring is large ($r=\SI{50}{m}$). The proton momentum has to be
$p=\SI{700.7}{MeV/c}$ in this case.

Another option indicated by the red star in Fig.~\ref{fig:p_r_proton} is
to operate a ring with a combined electric and magnetic field.
Using a field strength of $E=\SI{7}{MeV/m}$ and a moderate radial magnetic field of $B=\SI{0.03}{T}$. In this case
the proton momentum is $p \approx \SI{300}{MeV/c}$.
For this so called prototype ring the bending radius is only $r=\SI{9}{m}$.
In order to operate the beams clockwise and counter clockwise, the
magnetic field has to be reversed.
The CPEDM (charged particle EDM) collaboration is pursuing the design of such a so called prototype ring.
More details are discussed in the conference contributions~\cite{Javakhishvili,shankar,siddique}
and reference~\cite{Abusaif:2654645}.

\section{Summary and Conclusions}

Electric Dipole Moments are a unique probe to search fot CP-violating
interactions outside the Standard Model.
To directly measure EDMs of charged particles, like protons, deuterons
or muons, storage rings are needed. First measurements of the deuteron
EDM have been performed at the Cooler Synchrotron COSY at Forschungszentrum
J\"ulich, Germany.

To reduce systematic uncertainties a new type of storage ring
has to be constructed which allows one to operate clockwise and counter-clockwise
circulating beams in the so called frozen spin condition, where the spin
precession due to the magnetic moment is suppressed. Plans for such a ring were presented.

This work was supported by the ERC Advanced Grant (srEDM \#694340)
of the European Union.
I am grateful to Vera Shmakova and Swathi Karanth for preparing the plots for figure~\ref{fig:alpha_2},\ref{fig:map123} and \ref{fig:axion_sen}.








\bibliography{/home/pretz/bibtex/literature_axion.bib,/home/pretz/bibtex/literature_edm.bib,proceedings.bib}

\bibliographystyle{unsrt}


\end{document}